# The third exactly solvable hypergeometric quantum-mechanical potential

A.M. Ishkhanyan[1,2]

[1]Institute for Physical Research, NAS of Armenia, Ashtarak 0203, Armenia
[2]Institute of Physics and Technology, National Research Tomsk Polytechnic University, Tomsk 634050, Russia

We introduce the third independent exactly solvable hypergeometric potential, after the Eckart and the Pöschl-Teller potentials, which is proportional to an energy-independent parameter and has a shape that is independent of this parameter. The general solution of the Schrödinger equation for this potential is written through fundamental solutions each of which presents an irreducible combination of two Gauss hypergeometric functions. The potential is an asymmetric step-barrier with variable height and steepness. Discussing the transmission above such a barrier, we derive a compact formula for the reflection coefficient.

## 1. Introduction

Though much insight in quantum mechanics is gained from understanding the solutions of the Schrödinger equation in terms of special functions, such solutions are rare. Among these, of particular interest are the ones for which the involved parameters can be varied independently because owing to the complete analytic examination of the whole variation range of the possible physical effects covered by such potentials this option adds substantial generality and universality to the resulting observations.

However, perhaps somewhat unexpectedly, the number of such potentials is very limited. Besides, there are only a few such potentials that are independent in the sense that they cannot be derived from each other by specifications of the involved parameters. For instance, there is known only a handful of such independent potentials solvable in terms of the most known and most used set of the special functions - the functions of the hypergeometric class. Indeed, only three such potentials solvable in terms of the Kummer confluent hypergeometric functions have been known until the last year. These were the harmonic oscillator (plus inverse square) [1], the Coulomb (plus inverse square) potentials discussed by Schrödinger [1,2] and Kratzer [3], and the Morse potential [4]). Extending the collection to five members, we have recently presented two more confluent hypergeometric potentials - the inverse square root [5] and the Lambert-W step-barrier potentials [6].

The list of the potentials that are solved in terms of the Gauss hypergeometric functions, which are more general functions that involve one more variable parameter compared with the confluent hypergeometric functions, is interestingly even shorter - the list includes just two names - the Eckart [7] and the Pöschl-Teller [8] potentials. The widely discussed in the past Rosen-Morse [9], Manning-Rosen [10], Hulthén [11], Woods-Saxon [12], Scarf [13] and several other known ones are particular cases of these two potentials.

In the present paper we introduce the third exactly solvable Gauss hypergeometric potential. We present the general solution of the problem which is achieved by reduction of the Schrödinger equation to the general Heun equation [14-16] and further expansion of the solution of this equation in terms of the Gauss hypergeometric functions (see [17-20]). A peculiarity of the solution is that each of the two fundamental solutions that compose the general solution of the problem is given by an irreducible linear combination of two Gauss hypergeometric functions. We note that this feature, the two-term structure of each of the fundamental solutions of the problem, is common for all the three new exactly solvable (confluent and ordinary) hypergeometric potentials as well as several recently reported conditionally exactly solvable potentials [21-23].

The potential we introduce is defined on the whole coordinate axes. It is an asymmetric step-barrier the height and the steepness of which are controlled by two independent parameters. The potential involves two more independent parameters which stand for the energy origin and the position of the step. Discussing the above-barrier transmission, we derive an exact formula for the quantum-mechanical reflection coefficient.

**2. The potential and the solution**

The potential we consider is

$$V(x) = V_0 + \frac{V_1}{\sqrt{1+e^{2(x-x_0)/\sigma}}}. \quad (1)$$

This is an asymmetric step-barrier of height $V_1$ the steepness of which is controlled by the parameter $\sigma$ (figure 1). In the limit $\sigma \to 0$ the potential turns into the abrupt-step potential

$$V_{SP} = V_0 \text{ if } x < x_0 \text{ and } V_{SP} = V_0 + V_1 \text{ if } x > x_0. \quad (2)$$

The potential (1) has four independent parameters, $V_0, V_1, x_0$ and $\sigma$, which stand for the energy origin, the step height, the position of the step and the scaling of the space coordinate, respectively. We note that the sub-family of the potentials (1) generated by variation of $\sigma$ at fixed $V_{0,1}$ has a fixed point located at $x = x_0$: $V(x_0) = V_0 + V_1/\sqrt{2}$ (figure 1).

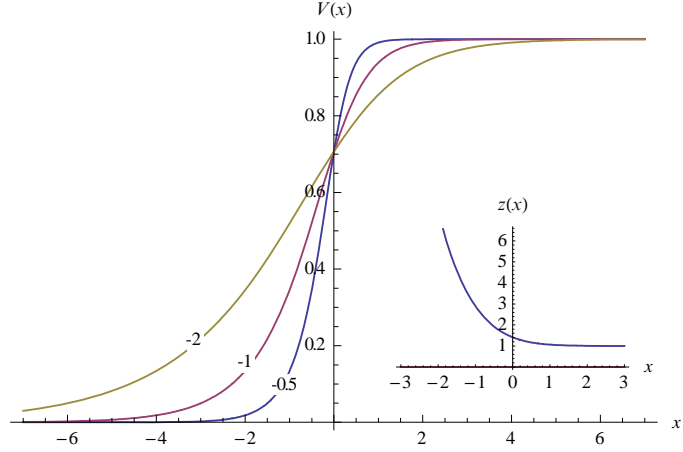

Fig.1. Potential (1) for $V_0 = 0$, $V_1 = 1$, $x_0 = 0$, $\sigma = -0.5, -1, -2$ ((in units $\hbar = m = 1$).
The inset presents the coordinate transformation $z(x)$.

It is straightforwardly checked that the one-dimensional stationary Schrödinger equation for a particle of mass $m$ and energy $E$:

$$\frac{d^2\psi}{dx^2} + \frac{2m}{\hbar^2}(E - V(x))\psi = 0, \quad (3)$$

for this potential with arbitrary (real or complex) parameters $V_{0,1}$ and $x_0, \sigma$ admits a fundamental solution written as

$$\psi(x) = (z+1)^{\alpha_1}(z-1)^{\alpha_2} u(z), \quad z = \sqrt{1 + e^{2(x-x_0)/\sigma}}, \quad (4)$$

$$u(z) = {}_2F_1\left(\alpha - 1, \beta; \gamma - 1; \frac{z+1}{2}\right) + \frac{(\alpha_2 - \alpha_1 + \beta z)}{2(\gamma - 1)} {}_2F_1\left(\alpha, \beta + 1; \gamma; \frac{z+1}{2}\right), \quad (5)$$

where the involved parameters are given as

$$(\alpha, \beta, \gamma) = (\alpha_1 + \alpha_2 - \alpha_0, \alpha_1 + \alpha_2 + \alpha_0, 1 + 2\alpha_1), \quad (6)$$

$$\alpha_{0,1,2} = \left(\pm\sqrt{\frac{-2m\sigma^2}{\hbar^2}(E - V_0)}, \pm\sqrt{\frac{-m\sigma^2}{2\hbar^2}(E - V_0 + V_1)}, \pm\sqrt{\frac{-m\sigma^2}{2\hbar^2}(E - V_0 - V_1)}\right). \quad (7)$$

Here any combination for the signs of $\alpha_{0,1,2}$ is applicable. We note that by choosing different combinations one can construct different fundamental solutions.

The potential and the corresponding solution are derived if one considers the reduction of the Schrödinger equation to the general Heun equation [14-16]

$$u_{zz} + \left(\frac{\gamma}{z - a_1} + \frac{\delta}{z - a_2} + \frac{\varepsilon}{z - a_3}\right)u_z + \frac{\alpha\beta z - q}{(z - a_1)(z - a_2)(z - a_3)}u = 0. \quad (8)$$

The technique for this reduction is based on the results of [24] and follows the particular lines developed for the quantum two-state problem in [25-27]. To avoid the overlap with the texts of these papers we just briefly outline the general derivation lines which are as follows.

The transformation of the variables $\psi = \varphi(z) u(z)$, $z = z(x)$ with

$$\varphi(z) = \rho(z)^{-1/2} \exp\left(\frac{1}{2}\int f(z)dz\right), \quad \rho(z) = dz/dx \quad (9)$$

reduces the Schrödinger equation (3) to the equation

$$I(z) = g - \frac{f_z}{2} - \frac{f^2}{4} = -\frac{1}{2}\left(\frac{\rho_z}{\rho}\right)_z - \frac{1}{4}\left(\frac{\rho_z}{\rho}\right)^2 + \frac{2m}{\hbar^2}\frac{E-V(z)}{\rho^2}, \quad (10)$$

where $f(z)$ and $g(z)$ are the coefficients of the general Heun equation (8) and $I(z)$ is the *invariant* [28] of that equation if it is rewritten in the Liouville *normal form* [29].

The basic assertion of [24] is that if a potential is proportional to an energy-independent parameter and has a shape which is independent of the energy and that parameter, then the logarithmic $z$-derivative $\rho'(z)/\rho(z)$ cannot have poles other than the finite singularities of the target equation to which the Schrödinger equation is reduced (see theorem (27) of [24]). It then follows that because the Heun equation has three finite singular points (located at $z = a_1, a_2, a_3$) the appropriate coordinate transformation is of the form

$$\rho = (z-a_1)^{m_1}(z-a_2)^{m_2}(z-a_3)^{m_3}/\sigma, \quad (11)$$

where $m_{1,2,3}$ are integers or half-integers and $\sigma$ is an arbitrary scaling constant.

Since the invariant $I(z)$ of the Heun equation (8) is a fourth-degree polynomial in $z$ divided by $(z-a_1)^2(z-a_2)^2(z-a_3)^2$, the next step is now to match the $\rho_z/\rho$-dependent terms of equation (10) with this form of the invariant. This leads to eleven independent cases which cover all known potentials for which the Schrödinger equation is solved in terms of hypergeometric functions and suggests several new developments [30]. The third exactly solvable hypergeometric potential that we have presented above is derived if one checks the solutions which are written in terms of a combination of two hypergeometric functions. If $a_{1,2,3} = (-1,1,0)$, the conditions for this to happen are $\varepsilon = -1$ and $q(q+\gamma-\delta) = \alpha\beta$ (see the details in [19]). It is then readily checked that these conditions are satisfied if $m_{1,2,3} = (1,1,-1)$. According to equation (9), we now put

$$\varphi(z) = (z-a_1)^{\alpha_1}(z-a_2)^{\alpha_2}(z-a_3)^{\alpha_3}, \quad (12)$$

and require, as the form of the invariant $I(z)$ indicates,

$$(z-a_1)^2(z-a_2)^2(z-a_3)^2 V(z)/\rho^2 = v_0 + v_1 z + v_2 z^2 + v_3 z^3 + v_4 z^4. \tag{13}$$

Finally, demanding the constants $v_{0,1,2,3,4}$ to be independent, we are straightforwardly led to the potential (1) and the corresponding solution of the Schrödinger equation (4)-(7).

## 3. Quantum-mechanical reflection at above-barrier transmission

The applications of the Schrödinger equation in contemporary physics cover an extremely wide set of effects in different branches. For this reason one may envisage many discussions of the presented potential. As an example we consider the quantum-mechanical reflection at transmission of a particle above this potential barrier.

We note that the coordinate transformation

$$z = \sqrt{1 + e^{2(x-x_0)/\sigma}} \tag{14}$$

maps the axes $x \in (-\infty, +\infty)$ into the interval $z \in (+\infty, 1)$ (see the inset of figure 1). It is then convenient to rewrite the general solution of the problem in the following equivalent form:

$$\psi(x) = (z+1)^{\alpha_1}(z-1)^{\alpha_2}\left(F(z) + \frac{\alpha_2 - \alpha_1 + bz}{ab}\frac{dF(z)}{dz}\right), \tag{15}$$

with

$$F(z) = C_1 \cdot {}_2F_1\left(a,b;c;\frac{1+z}{2}\right) + C_2 \cdot {}_2F_1\left(a,b;a+b-c+1;\frac{1-z}{2}\right), \tag{16}$$

where $C_{1,2}$ are arbitrary constants and the parameters involved in the hypergeometric functions are given as $(a,b,c) = (\alpha_1 + \alpha_2 - \alpha_0 - 1, \alpha_1 + \alpha_2 + \alpha_0, 2\alpha_1)$ (compare with equation (6)). The advantage of this choice is that demanding the wave function at $x \to +\infty$ to involve only one plane wave, that is only the transmitted wave, we get that $C_1 = 0$ and

$$\psi(+\infty) \sim C e^{ik_2 x}, \quad C = 2^{\alpha_1 - \alpha_2 - 2}\left(2 - \frac{\alpha_0}{\alpha_2}\right)C_2, \quad k_2 = \sqrt{\frac{2m}{\hbar^2}(E - V_0 - V_1)}. \tag{17}$$

Expanding now the solution at $x \to -\infty$, we get the asymptote

$$\psi(-\infty) \sim A e^{+ik_1 x} + B e^{-ik_1 x}, \quad k_1 = \sqrt{\frac{2m}{\hbar^2}(E - V_0)} \tag{18}$$

with

$$A = -\frac{2^{\alpha_1+\alpha_2-\alpha_0}(\alpha_1-\alpha_2)(\alpha_1+\alpha_2)\Gamma(2\alpha_0)\Gamma(2\alpha_2)}{\Gamma(\alpha_0-\alpha_1+\alpha_2+1)\Gamma(\alpha_0+\alpha_1+\alpha_2+1)}C_2, \tag{19}$$

$$B = \frac{2^{\alpha_1+\alpha_2+\alpha_0}\Gamma(-2\alpha_0)\Gamma(2\alpha_2)}{\Gamma(-\alpha_0-\alpha_1+\alpha_2)\Gamma(-\alpha_0+\alpha_1+\alpha_2)}C_2, \tag{20}$$

where $\Gamma$ is the Euler gamma-function. The transmission coefficient is then determined as

$$T = \left|\frac{k_2 C^2}{k_1 A^2}\right| = \frac{2\sinh(2\pi\sigma k_1)\sinh(\pi\sigma k_2)}{\cosh[\pi\sigma(2k_1+k_2)] - \cosh[\pi\sigma\sqrt{2k_1^2 - k_2^2}]}. \quad (21)$$

As expected, in the limit $\sigma \to 0$ this recovers the result for the abrupt-step potential [31]:

$$T = T_{SP} + O(\sigma^2), \quad T_{SP} = \frac{4k_1 k_2}{(k_1 + k_2)^2}. \quad (22)$$

It is readily checked that the correction term is always positive so that we conclude that because of the smoothness the transmission above the potential (1) is always more than that for the abrupt-step potential (2). In the infinitely-smooth limit $\sigma \to \infty$ we have $T|_{\sigma \to \infty} \to 1$ so that in this idealized limit the potential becomes transparent. The reflection coefficient $R = 1 - T$ as a function of the energy is shown in figure 2.

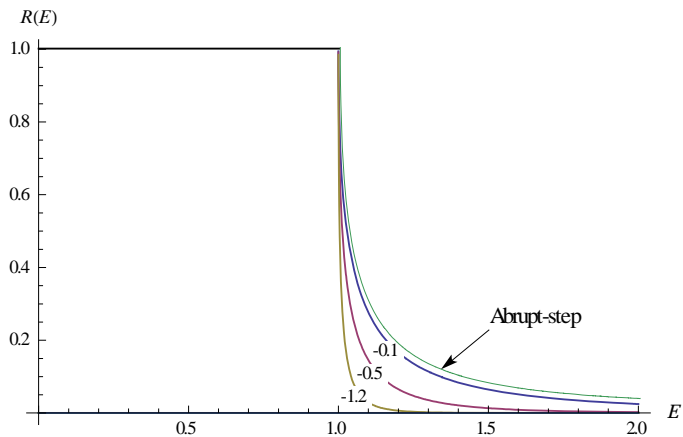

Fig.2. The reflection coefficient $R = 1-T$ versus energy $E$ for $V_0 = 0$, $V_1 = 1$, $x_0 = 0$, $\sigma = -0.1, -0.5, -1.2$ (in units $\hbar = m = 1$).

## 4. Discussion

Thus, we have presented one more independent exactly solvable hypergeometric potential for which all the involved parameters are varied independently. This is a four-parametric asymmetric step-barrier potential with controlled height and steepness. We have discussed the quantum reflection at transmission of a particle above this potential and have derived a compact formula for the reflection coefficient.

The potential is derived via reduction of the Schrödinger equation to the general Heun equation [14] which is a natural generalisation of the Gauss hypergeometric equation.

Though this is a quite complicated mathematical object (in that it has an additional singularity as compared with the Gauss hypergeometric equation and it involves an accessory parameter that is absent in the hypergeometric case) and its analytic theory currently needs much development, in several cases the solution can be expressed in terms of simpler mathematical functions via direct Heun-to-hypergeometric reductions by a variable transformation (see, e.g., [32,33]). A different approach had been put forward by Svartholm and Erdélyi who proposed series expansions of the Heun functions using the hypergeometric functions as expansion functions [17,18]. We have recently developed several other such expansions involving both ordinary and generalised hypergeometric functions [19,34].

These expansions open a possibility to construct closed-form solutions by termination of the series. In the present paper we have used a particular two-term termination discussed, e.g., in [19]. The result is the presented potential for which the general solution of the Schrödinger equation is written through fundamental solutions that are linear combinations of two Gauss hypergeometric functions. It seems that this employment of the series solutions of the Heun equations is rather useful [35] and can be applied to other structurally similar problems, for instance, to the Dirac and Klein-Gordon equations [36,37].

We would like to conclude that from the point of view of the super-symmetric quantum mechanics [38,39] the presented potential is the superpotential of the partner potentials recently reported by A. López-Ortega [23].

**Acknowledgments**

This research has been conducted within the scope of the International Associated Laboratory IRMAS (CNRS-France & SCS-Armenia). The work has been supported by the Armenian State Committee of Science (SCS Grants No. 13RB-052 and No. 15T-1C323) and the project "Leading Russian Research Universities" (Grant No. FTI_120_2014 of the Tomsk Polytechnic University).

**References**
1. E. Schrödinger, "Quantisierung als Eigenwertproblem (Erste Mitteilung)", Annalen der Physik **76**, 361-376 (1926).
2. E. Schrödinger, "Quantisierung als Eigenwertproblem (Zweite Mitteilung)". Annalen der Physik **79**, 489-527 (1926).
3. A. Kratzer, "Die ultraroten Rotationsspektren der Halogenwasserstoffe", Z. Phys. **3**, 289-307 (1920).
4. P.M. Morse, "Diatomic molecules according to the wave mechanics. II. Vibrational levels", Phys. Rev. **34**, 57-64 (1929).

5.  A.M. Ishkhanyan, "Exact solution of the Schrödinger equation for the inverse square root potential $V_0/\sqrt{x}$ ", EPL **112**, 10006 (2015).
6.  A.M. Ishkhanyan, "The Lambert W-barrier - an exactly solvable confluent hypergeometric potential", Phys. Lett. A **380**, 640-644 (2016).
7.  C. Eckart, "The penetration of a potential barrier by electrons", Phys. Rev. **35**, 1303-1309 (1930).
8.  G. Pöschl, E. Teller, "Bemerkungen zur Quantenmechanik des anharmonischen Oszillators", Z. Phys. **83**, 143-151 (1933).
9.  N. Rosen and P.M. Morse, "On the vibrations of polyatomic molecules", Phys. Rev. **42**, 210-217 (1932).
10. M.F. Manning and N. Rosen, "A potential function for the vibrations of diatomic molecules", Phys. Rev. **44**, 953-953 (1933).
11. L. Hulthén, Ark. Mat. Astron. Fys. **28A**, 5 (1942); L. Hulthén, "Über die Eigenlösungen der Schrödinger-Gleichung der Deuterons", Ark. Mat. Astron. Fys. **29B**, 1-12 (1942).
12. R.D. Woods and D.S. Saxon, "Diffuse surface optical model for nucleon-nuclei scattering", Phys. Rev. **95**, 577-578 (1954).
13. F. Scarf, "New soluble energy band problem", Phys. Rev. **112**, 1137-1140 (1958).
14. K. Heun, "Zur Theorie der Riemann'schen Functionen Zweiter Ordnung mit Verzweigungspunkten", Math. Ann. **33**, 161 (1889).
15. A. Ronveaux (ed.), *Heun's Differential Equations* (Oxford University Press, London, 1995).
16. S.Yu. Slavyanov and W. Lay, *Special functions* (Oxford University Press, Oxford, 2000).
17. N. Svartholm, "Die Lösung der Fuchs'schen Differentialgleichung zweiter Ordnung durch Hypergeometrische Polynome", Math. Ann. **116**, 413 (1939).
18. A. Erdélyi, "Certain expansions of solutions of the Heun equation", Q. J. Math. (Oxford) **15**, 62 (1944).
19. T.A. Ishkhanyan, T.A. Shahverdyan, A.M. Ishkhanyan, "Hypergeometric expansions of the solutions of the general Heun equation governed by two-term recurrence relations for expansion coefficients", arXiv:1403.7863 [math.CA] (2016).
20. T.A. Ishkhanyan and A.M. Ishkhanyan, "Expansions of the solutions to the confluent Heun equation in terms of the Kummer confluent hypergeometric functions", AIP Advances **4**, 087132 (2014).
21. A. López-Ortega, "New conditionally exactly solvable inverse power law potentials", Phys. Scr. **90**, 085202 (2015).
22. A.M. Ishkhanyan, "A conditionally exactly solvable generalization of the inverse square root potential", arXiv:1511.03565 (2016).
23. A. López-Ortega, "A conditionally exactly solvable generalization of the potential step", arXiv:1512.04196 [math-ph] (2015); A. López-Ortega, "New conditionally exactly solvable potentials of exponential type", arXiv:1602.00405 [math-ph] (2016).
24. A. Ishkhanyan and V. Krainov, "Discretization of Natanzon potentials", arXiv:1508.06989 [quant-ph] (2015).
25. T.A. Shahverdyan, T.A. Ishkhanyan, A.E. Grigoryan, A.M. Ishkhanyan, "Analytic solutions of the quantum two-state problem in terms of the double, bi- and triconfluent Heun functions", J. Contemp. Physics (Armenian Ac. Sci.) **50**, 211-226 (2015).
26. A.M. Ishkhanyan and A.E. Grigoryan, "Fifteen classes of solutions of the quantum two-state problem in terms of the confluent Heun function", J. Phys. A **47**, 465205 (2014).
27. A.M. Ishkhanyan, T.A. Shahverdyan, T.A. Ishkhanyan, "Thirty five classes of solutions of the quantum time-dependent two-state problem in terms of the general Heun functions", Eur. Phys. J. D **69**, 10 (2015).
28. E.L. Ince, *Ordinary Differential Equations* (Dover Publications, New York, 1964).

29. J. Liouville, "Second mémoire sur le développement des fonctions ou parties de fonctions en séries dont divers termes sont assujettis à satisfaire a une même équation différentielle du second ordre contenant un paramètre variable", J. Math. Pures Appl. **2**, 16-35 (1837).
30. A.M. Ishkhanyan, "Schrödinger potentials solvable in terms of the general Heun functions", arXiv:1601.03360 [quant-ph] (2016).
31. S. Flügge, *Practical Quantum Mechanics I, II* (Springer Verlag, Berlin, 1971).
32. R.S. Maier, "On reducing the Heun equation to the hypergeometric equation", J. Diff. Equations **213**, 171 (2005).
33. R. Vidunas, G. Filipuk, "Parametric transformations between the Heun and Gauss hypergeometric functions", Funkcialaj Ekvacioj **56**, 271 (2013).
34. A.M. Ishkhanyan, "The Appell hypergeometric expansions of the solutions of the general Heun equation", arXiv:1405.2871 [math-ph] (2016).
35. A.M. Ishkhanyan, "Schrödinger potentials solvable in terms of the confluent Heun functions", Theor. Math. Phys., in press (2016).
36. A.S. Tarloyan, T.A. Ishkhanyan, and A.M. Ishkhanyan, "Four five-parametric and five four-parametric independent confluent Heun potentials for the stationary Klein-Gordon equation", Ann. Phys. doi: 10.1002/andp.201500314 (2016).
37. *Heun functions, their generalizations and applications*, http://theheunproject.org/bibliography.html
38. A. Gangopadhyaya, J.V. Mallow, C. Rasinariu, *Supersymmetric quantum mechanics: An introduction* (World Scientific, Singapore, 2010).
39. B. Bagchi, *Supersymmetry in Quantum and Classical Physics* (Chapman and Hall/CRC Press, Boca Raton, 2000).